\documentclass{caosp309}

\usepackage{graphicx}

\usepackage{natbib}
\articleNo{301}
\pubyear{2020}
\volume{50}
\volnumber{3}
\firstpage{1}
\received{May 1, 2020}
\accepted{July 28, 2020}

\def\BibTeX{{\rm B\kern-.05em{\sc i\kern-.025em b}\kern-.08em
             T\kern-.1667em\lower.7ex\hbox{E}\kern-.125emX}}

\begin{document}

%
\hauthor{A. Liakos}

\title{The catalogue of $\delta$~Sct pulsators in binary systems in 2024}


%
%
\author{        A.\,Liakos\orcid{0000-0002-0490-1469}       }

\institute{Institute for Astronomy, Astrophysics, Space Applications and Remote Sensing, National Observatory of Athens, Metaxa \& Vas. Pavlou St., GR-15236, Athens, Greece \email{alliakos@noa.gr}}

\date{March 8, 2024}

\maketitle

\begin{abstract}
The updated catalogue of $\delta$~Sct stars in binary systems as well as their statistical properties are presented. Thanks to the Kepler, K2 and TESS space missions, this version of the catalogue contains more than 1000 $\delta$~Sct pulsators in binaries. The sample is divided according to the Roche Geometry of the binary systems in order to check for any systematic differences in the pulsators' evolution due to the proximity of the companion star. Statistics, demographics, and distributions of these pulsating stars within the H-R and mass-radius diagrams are provided. We notice that the absolute parameters have been accurately determined for only approximately 10\% of the whole sample. Additionally, updated correlations between pulsation and orbital periods and evolutionary status are presented. This work aims to motivate the researchers for systematic analyses of such objects in order to increase the sample of systems with well known physical properties.
\end{abstract}

%
\section{Introduction}
\label{Intro}
\label{intr}

$\delta$~Scuti stars (DS) are short-period pulsating variables that can oscillate in both radial and non-radial modes. The radial and low-order non-radial modes are driven by the kappa mechanism \citep{AER10, BAL15}, while the higher-order non-radial modes result from turbulent pressure in the hydrogen convective zone \citep{ANT14, GRA15}. These stars typically have masses between 1.4 and 2.5~M$_{\odot}$, temperatures ranging from approximately 6000 to 9500~K, and belong to luminosity classes III-V. Most of them are found within the classical instability strip \citep{MUR19}.

Although the first DS-members of binary systems were discovered in '70s \citep[e.g.][]{HUD71}, since the early '00s there was not any systematic listing of them, mainly due to the limited discoveries. DS in binaries were difficult to be detected before 2000 mostly because the photometric accuracy of photometers and first generations CCDs was lower than the amplitudes of their oscillations. The first list of these stars (eight in total) was published by \citet{MKR02}, who introduced the term `Oscillating Eclipsing Algols' (oEA stars) for semidetached eclipsing systems with a mass accretor DS component. Later on, \citet{MKR04, MKR05} announced a few more similar systems, while \citet{SOY06a, SOY06b} published 25 cases expanding also to the detached type systems. The latter works proposed also a correlation between the orbital ($P_{\rm orb}$) and pulsation ($P_{\rm pul}$) periods. \citet[][hereafter paper~I]{LIA12a} increased the sample of these systems to 74 and established the $P_{\rm orb}$-$P_{\rm pul}$ relation. The boundary between the end of the `dark ages' and the beginning of the `renaissance' era for this topic was the launch of the NASA/Kepler mission in 2009 \citep{KOC10}. After the first data release (DR1) of Kepler in 2014, the systematic discoveries of this kind of systems increased. \citet[][hereafter paper~II]{LIA17a} published an updated catalogue hosting 203 DS in 199 systems and found that there is a limit of 13~days in $P_{\rm orb}$ beyond that $P_{\rm orb}$ and $P_{\rm pul}$ are uncorrelated. After the end of Kepler and K2 missions, the NASA/TESS mission \citep{RIC15} was launched in 2018. A few months later the TESS-DR1 was announced, leading to hundreds of new discoveries. Correlations between the fundamental parameters of the stars of this kind of systems according to their Roche geometry (e.g.~$P_{\rm orb}$-$P_{\rm pul}$, $\log g$-$P_{\rm pul}$ correlations) were updated by \citet[][for semidetached systems]{LIA17b} and \citet[][for detached systems]{LIA20a}.

Due to the aforementioned space missions, hundreds of new systems of this kind have been discovered and to date (i.e.~August~2024) they have reached the number of 1048 DS in 1043 binary systems. However, there is still limited information on the absolute properties of their DS members. It should be noted that in ESA/Gaia~DR3 \citep{GAI23} there are a lot of radial velocities available for eclipsing binaries with a DS member, which it is expected to increase the number of DS with accurately determined physical properties. Therefore, after the scheduled end of TESS and the fourth data release (DR4) of Gaia in 2025, as well as the launch of the ESA/PLATO mission \citep{RAU16} in 2026, it is anticipated that this topic will enter through its `golden ages' regarding the exponential increase of the sample and the information on the physical properties of the DS.


\section{Demographics and statistical properties}
\label{Stat}

The sources of the present catalogue are: a)~the previous catalogues given in papers~I and II, b)~papers on satellite data mining and long lists of new cases \citep[e.g.][]{MUR18a, GAU19, KAH22c}, and c)~personal checks on the old rejected and ambiguous cases using satellite short-cadence data and new ground-based observations. Table~\ref{Tab:Dem1} contains the demographics of all DS-members of binary systems according to their type of variability and Roche geometry. In the same table, the respective numbers of DS, whose the physical properties have been determined, are also given. It should to be noted that to date only 66 DS out of 1048 belong to both eclipsing (E) and double-lined spectroscopic systems (SB2), which are considered as the utmost tools for determining the absolute properties. There are also another 32 DS with derived physical parameters that belong either to SB1+E or to SB1(2)+ellipsoidal systems and consist the second more reliable group for the physical parameters calculation. The absolute parameters of the rest 77 stars were determined using assumptions (e.g.~the mass of the primary was assumed based on its spectral type).


\begin{table}[t]
\small
\begin{center}
\caption{Demographics of DS in binary systems according to the type of variability and Roche geometry of their host systems. Values in parentheses denote the number of DS with determined mass and radius.}
\label{Tab:Dem1}
\begin{tabular}{lccc|c}
\hline\hline										
Variability type	&	Detached	&	Semidetached	&	Unclassified	&	All	\\
\hline									
SB2+E	&	35(31)	&	34(34)	&	5(1)	&	74(66)	\\
E and el	&	40(15)	&	53(32)	&	399(11)	&	492(58)	\\
SB1+E or SB1(2)+el	&	19(17)	&	15(12)	&	8(3)	&	42(32)	\\
SB1 and SB2	&	27(7)	&	0	&	11(5)	&	38(12)	\\
O--C and PM and Vis.	&	348(6)	&	0	&	54(1)	&	402(7)	\\
\hline									
Sum	&	469(76)	&	102(78)	&	477(21)	&	1048(175)	\\
\hline\hline		
\end{tabular}
\end{center}
SB1(2)=Single (Double) lined spectroscopic binary, E=Eclipsing binary, el=ellipsoidal variable, O$-$C=binarity detected through the periodic variation of the pulsational period, PM=binarity detected though the phase modulation of the pulsations, Vis.=Visual binary.
\small
\begin{center}
\caption{Demographics of DS in binary systems according to our knowledge on their dominant pulsational ($P_{\rm pul}$) and orbital ($P_{\rm orb}$) periods.}
\label{Tab:Dem2}
\begin{tabular}{lccc|c}
\hline\hline									
	&	$P_{\rm pul}$ \& $P_{\rm orb}$	&	Only $P_{\rm pul}$	&	Only $P_{\rm orb}$	&	All	\\
\hline	
Detached	&	200	&	5	&	264	&	469	\\
Semidetached	&	99	&	0	&	2	&	101	\\
Unclassified	&	350	&	59	&	69	&	478	\\
\hline									
Sum	&	649	&	64	&	335	&	1048	\\
\hline\hline									
\end{tabular}
\end{center}
\end{table}

Table~\ref{Tab:Dem2} lists the DS in binary systems according to their geometry and our knowledge on their $P_{\rm orb}$ and $P_{\rm pul}$. Systems with only the $P_{\rm pul}$ known are those for which DS-type pulsations have been detected and their duplicity has been confirmed, but there is no exact value of their $P_{\rm orb}$ (e.g.~wide orbit systems). On the other hand, those with only the $P_{\rm orb}$ known are those for which their $P_{\rm orb}$ has been determined, they present DS-type pulsations, but due to the limited time resolution of their data (e.g.~long-cadence data of Kepler, K2, and TESS), their dominant pulsation frequency may be higher than the nyquist frequency of these data sets (i.e.~$\sim 24$~d$^{-1}$).

\begin{figure}
\begin{tabular}{cc}
\includegraphics[width=6cm]{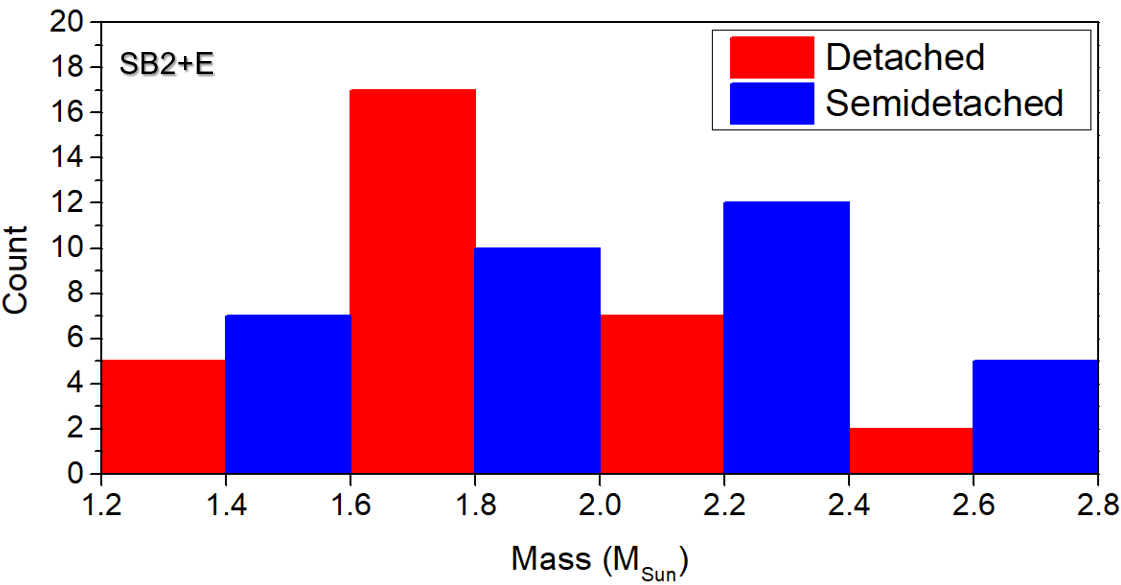}&\includegraphics[width=6cm]{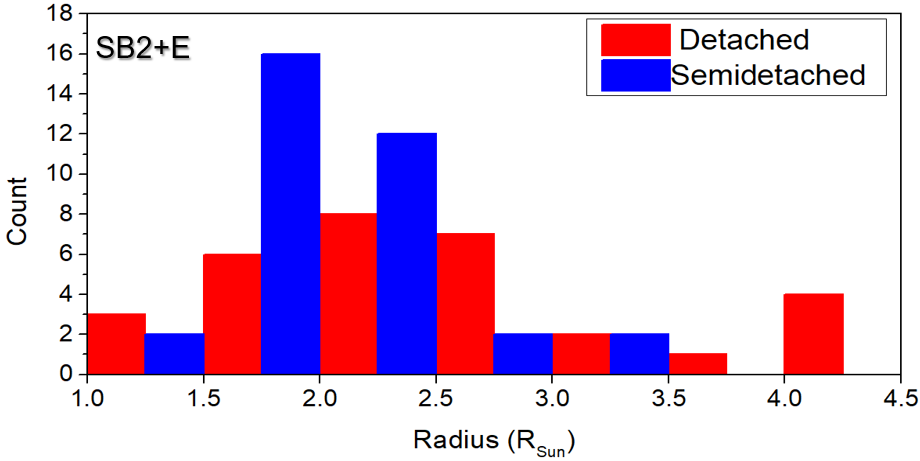}\\
\includegraphics[width=6cm]{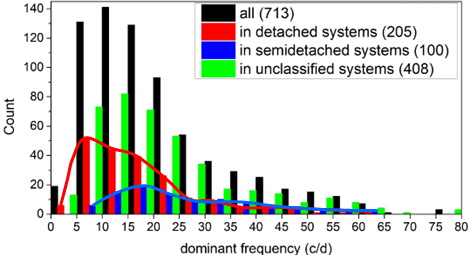}&\includegraphics[width=6cm]{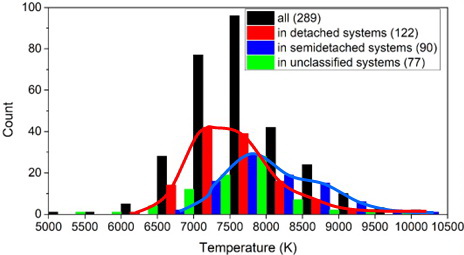}
\end{tabular}
\caption{Distributions of the: a) masses (upper left), b) radii (upper right), c) dominant frequencies (lower left), and d) temperatures (lower right) of the DS in binaries according to their Roche geometry.}
\label{Fig:Stat}
\end{figure}

The statistical properties of DS in binaries are illustrated in Fig.~\ref{Fig:Stat}. In this figure, for the masses and radii plots, we used only the DS that belong to SB2+E systems (Table~\ref{Tab:Dem1}). These plots show that DS in semidetached systems appear more compact. For the dominant frequencies and temperatures plots, we used 713 (Table~\ref{Tab:Dem2}) and 289 DS, respectively and we notice that the DS in detached systems exhibit slower pulsations and are slightly cooler in comparison with those in semidetached systems.

Using short- or medium- cadence data of the aforementioned space missions (i.e.~time resolutions 1-2~min and 10~min, respectively) and personal observations using the 1.2~m Kryoneri telescope \citep{XIL18} for $\sim25$ cases lacking space data, we examined in total 272 previously rejected and ambiguous cases of older catalogues \citep[][papers~I and II]{SOY06b}. Particularly, we searched the data of these stars/systems in order to detect either pulsations or eclipses. We found that 66 cases still remain ambiguous either for the pulsations exhibition or the pulsations type. For these systems, only long- and/or medium- cadence space data exist so far, which are not sufficient to determine with high certainty their potential pulsational properties. For another 40 cases, except for the eclipses search, we did not apply any other method to check whether a DS star of the list belongs to a binary system, so they remained as ambiguous regarding their duplicity. Finally, 166 cases were definitely rejected mostly due to the absence of pulsations.

\section{Correlations}
\label{Cor}

\begin{figure}
\centerline{\includegraphics[width=8.5cm]{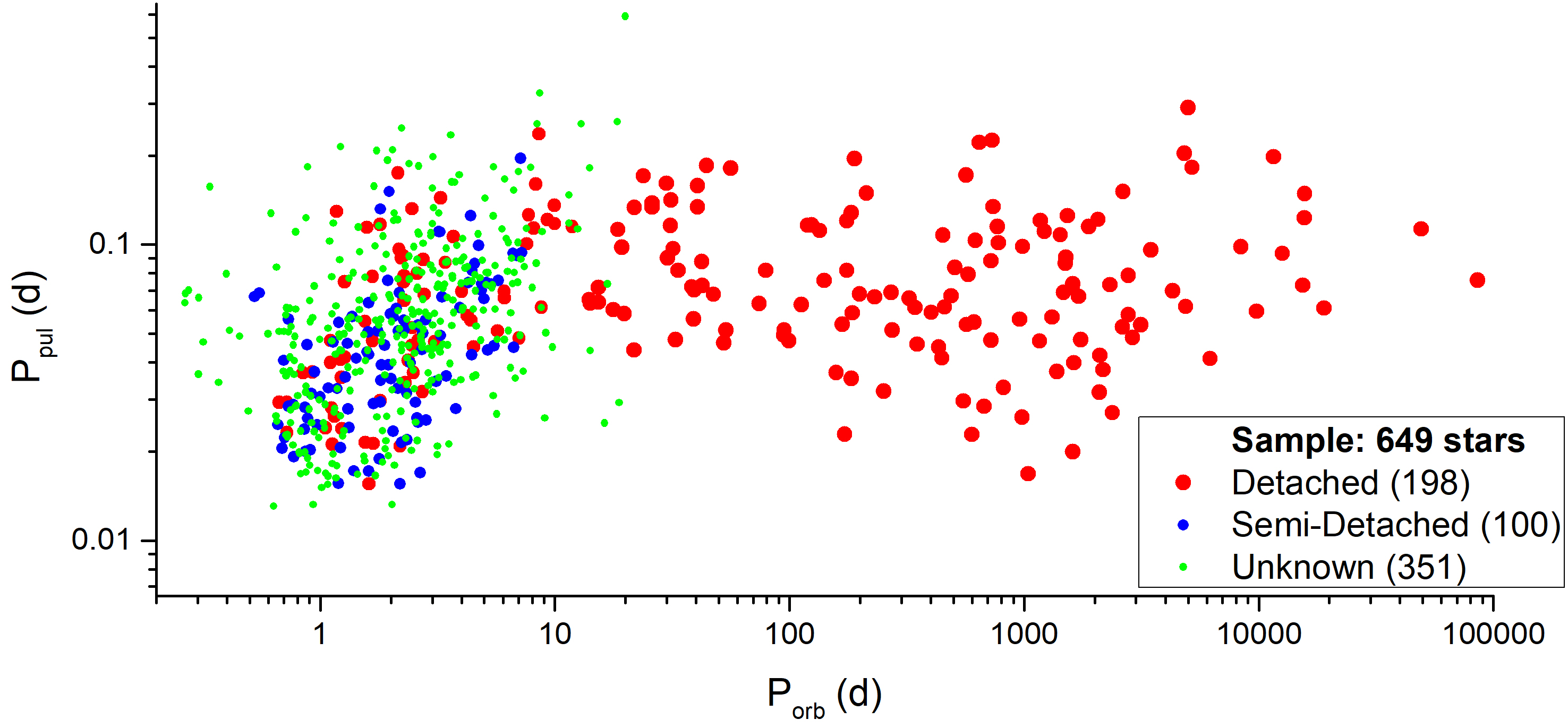}}
\caption{$P_{\rm orb}$-$P_{\rm pul}$ plot of the DS in binaries according to their Roche geometry.}
\label{Fig:PP_all}


\centerline{\includegraphics[width=5.7cm]{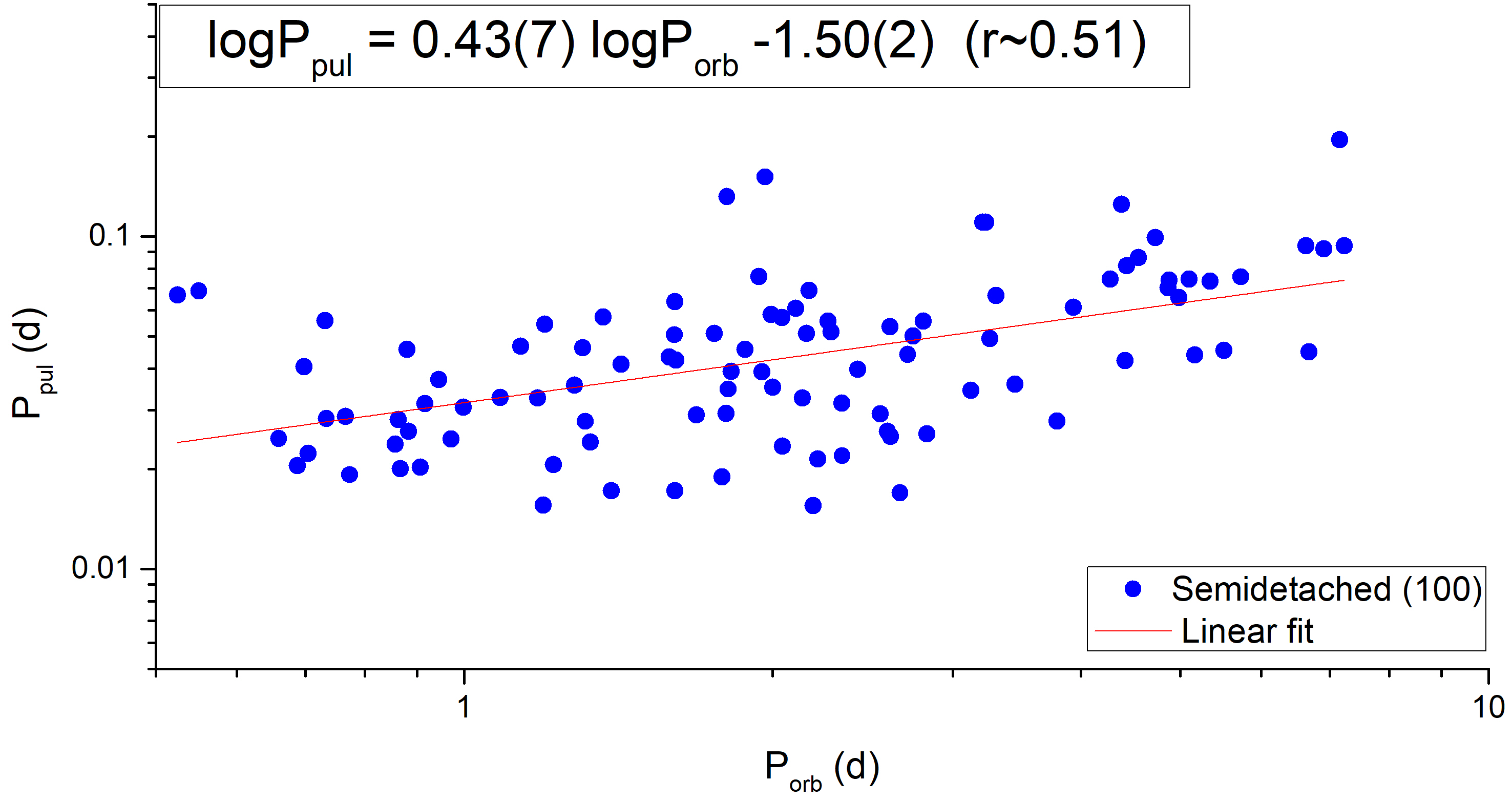}
\hspace{0.2cm}
\includegraphics[width=5.7cm]{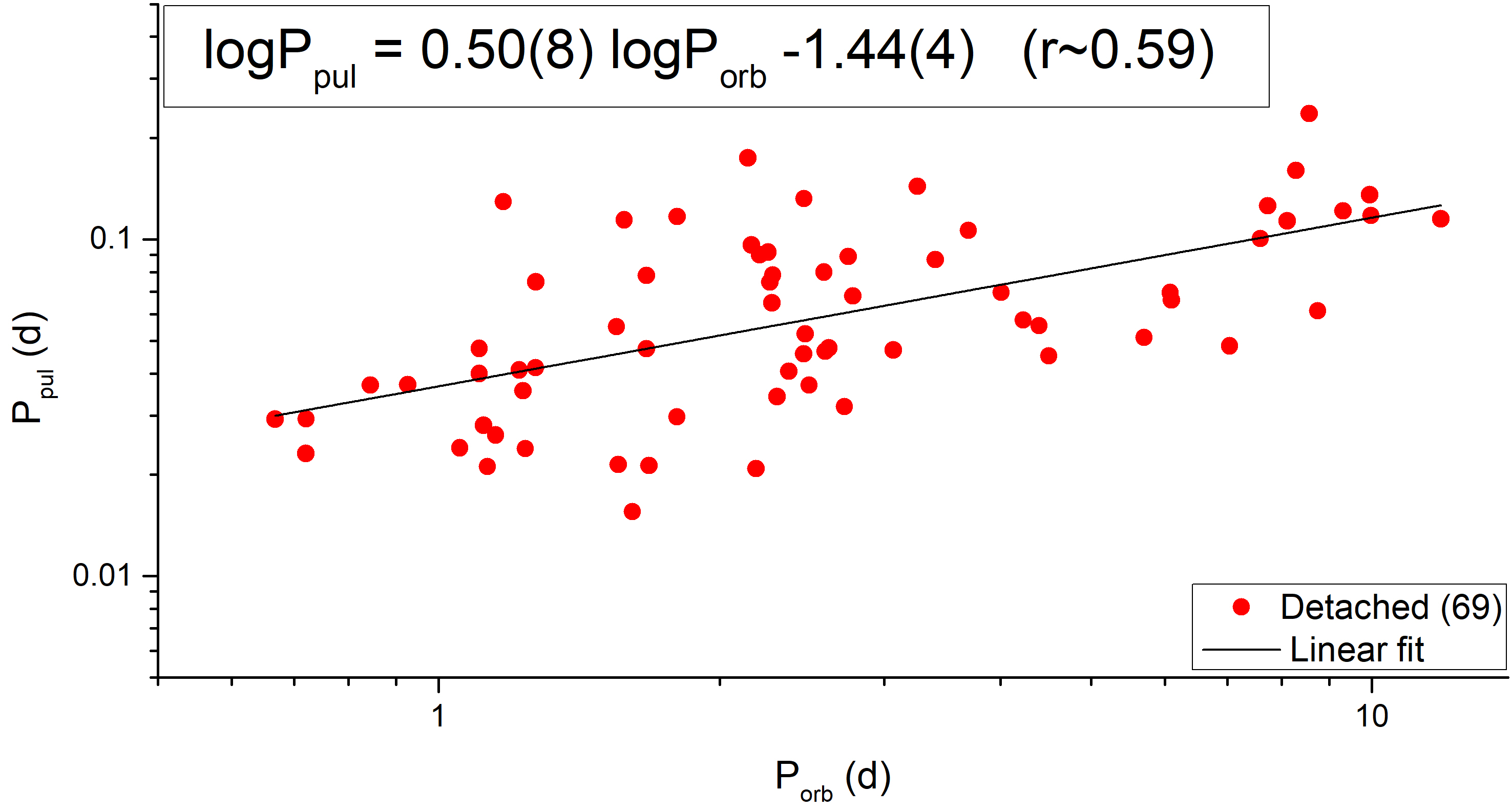}}
\caption{Linear fittings on the $P_{\rm orb}$-$P_{\rm pul}$ points of semidetached systems (left) and detached systems with $P_{\rm orb}<12.5$~d (right).}
\label{Fig:PPFIT}
\end{figure}

Using a much larger sample than that of the previous studies (i.e.~papers~I and II), we re-tested the correlation between $P_{\rm orb}$ and $P_{\rm pul}$ (Fig.~\ref{Fig:PP_all}). For this plot, systems that lack of modelling (i.e. previously categorized as unclassified) and have: a)~$P_{\rm orb}>20$~d and/or b)~eccentric orbits are classified directly as detached systems. Therefore, the remaining 351 unclassified systems in Fig.~\ref{Fig:PP_all} are short-period systems that lack of modelling.

Regarding the 198 detached systems, we applied the following methods to re-determine the threshold of $P_{\rm orb}$ beyond which $P_{\rm orb}$ and $P_{\rm pul}$ are uncorrelated. We fitted polynomial and exponential curves on the data points and by calculating their local extrema we found a limit of $P_{\rm orb}$=12.5~d and 13~d, respectively. Moreover, we performed linear fittings on datasets with various $P_{\rm orb}$ ranges (i.e.~up to 6, 10, 14, 18, 20, 26, 34, 40, and 45~days) and calculated the respective correlation coefficients ($r$). The two best correlations were found for the datasets with $P_{\rm orb}<10$~d and $P_{\rm orb}<14$~d, which is in very good agreement with the previous two methods. For semidetached systems, we just applied linear fitting on all data (100 systems), since there is no system with $P_{\rm orb}>7.5$~d within this subgroup. The fittings on the aforementioned datasets and the correlation formulae along with their respective $r$ are shown in Fig.~\ref{Fig:PPFIT}.

\begin{figure}[t]
\centerline{\includegraphics[width=7.5cm]{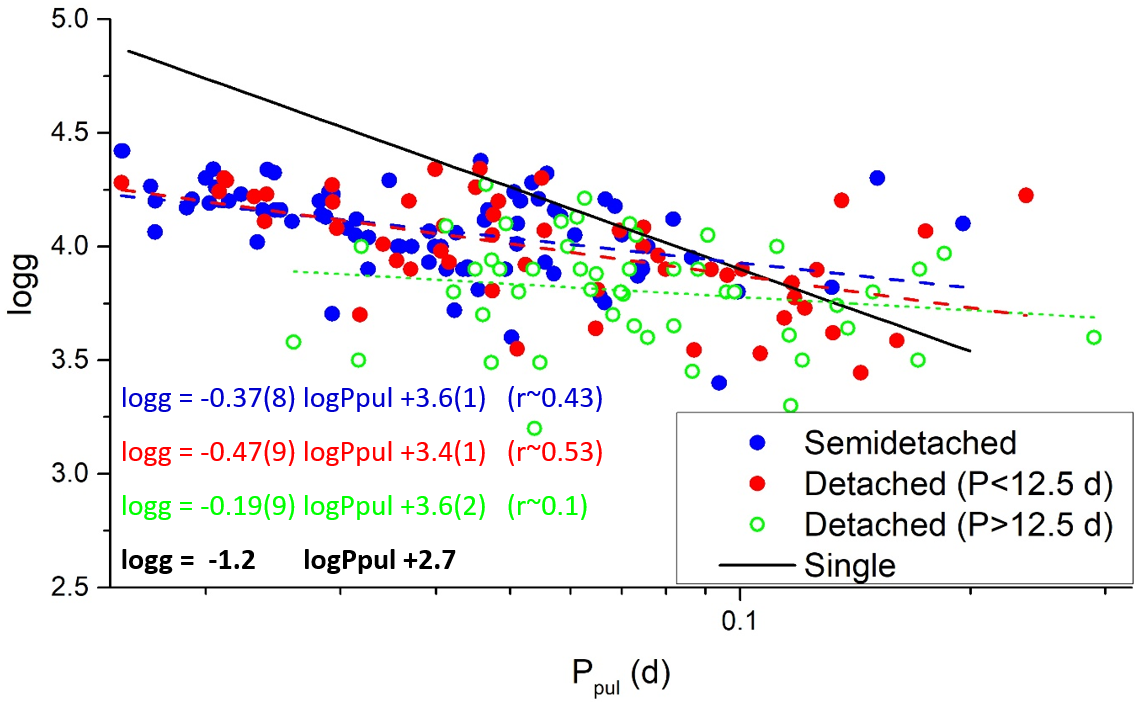}}
\caption{Correlations between $\log g$ and $P_{\rm pul}$ of binary DS according to their Roche geometry and $P_{\rm orb}$. The coloured lines are linear fittings on the different data sets (see text), while the black line, which represents the single DS, is given for comparison.}
\label{Fig:gP}
\end{figure}

\newpage
Another informative correlation is that between the evolutionary status ($\log g$) and the dominant $P_{\rm pul}$ of the binary DS. We performed linear fittings on the $\log g - P_{\rm pul}$ values (Fig.~\ref{Fig:gP}) of DS members of: a)~semidetached systems, b)~detached systems with $P_{\rm orb}<12.5$~d, and c)~detached systems with $P_{\rm orb}>12.5$~d. In contrast with our previous work (paper~II), the DS in semidetached and short-period detached systems have very similar relations (within the error limits) between $\log g - P_{\rm pul}$. On the other hand, very interestingly, for DS in detached systems with $P_{\rm orb}>12.5$~d, these quantities seem uncorrelated. In any case, the comparison between the relations of the DS in binaries with that of single DS (see Fig.~\ref{Fig:gP}) shows clearly that they follow different trend, which undoubtedly is related to the existence of the companion star. However, the most important in this plot is the fact the binary DS preserve their pulsations longer in comparison with the single DS.

\section{Evolutionary diagrams}
\label{Evol}

\begin{figure}[t]
\centerline{\includegraphics[width=6cm]{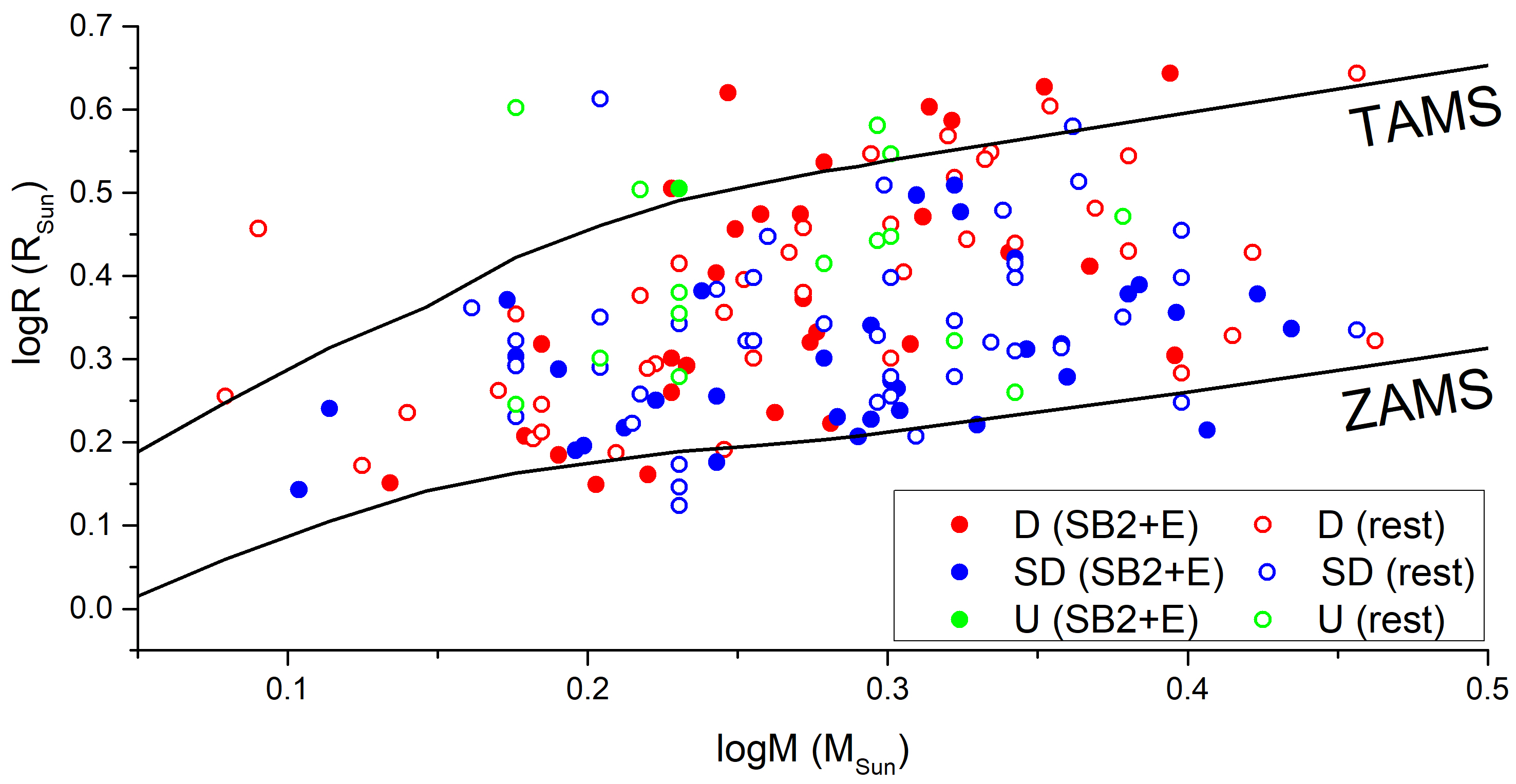}
\hspace{0.2cm}
\includegraphics[width=6cm]{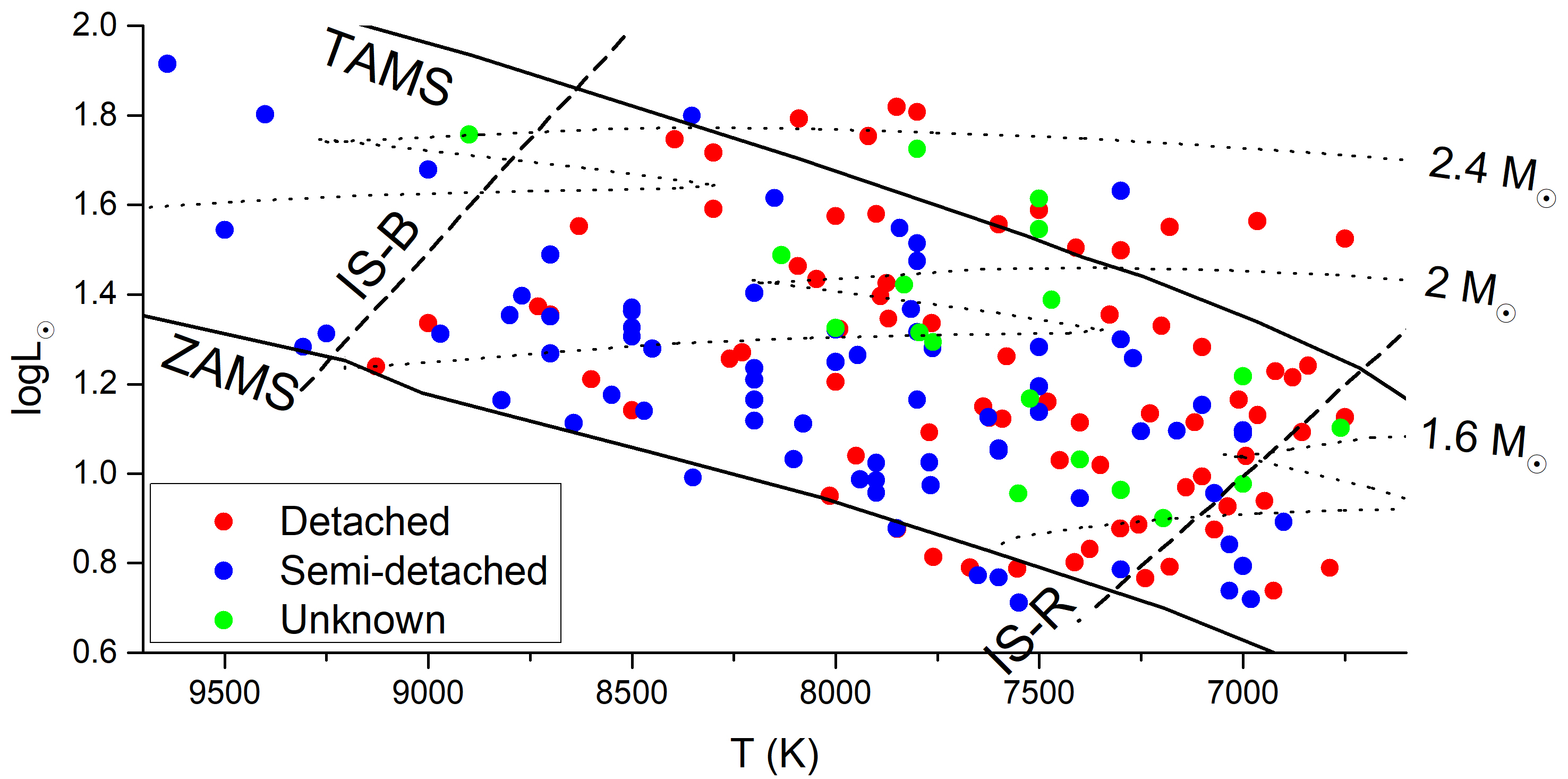}}
\caption{Mass-radius (left) and H--R (right) diagrams of DS in binaries.}
\label{Fig:Evol}
\end{figure}

Based on the new sample, we plot the DS of binary systems within the Mass-Radius and H--R diagrams in Fig.~\ref{Fig:Evol}. For the Mass-Radius diagram, we include all the DS with somehow determined absolute parameters (Table~\ref{Tab:Dem1}). In this plot, the DS that belong to SB2+E are denoted with different symbols. Regarding the H--R diagram, the sample was larger, since both temperature and luminosity can be more easily determined in comparison with the mass and the radius. Both diagrams clearly show that the vast majority of binary DS are Main-Sequence stars in contrast with the single DS \citep{UYT11}. Only a few cases have been found below the Zero-Age (ZAMS) or beyond the Terminal-Age (TAMS) Main-Sequence. Their vast majority lie within the classical instability strip (IS), and only a few cases are located beyond its blue limit (IS-B). On the contrary, there are many DS very close to the red boundary of IS (IS-R) and some of them are found beyond that. It should to be noted that this region of the H--R diagram has an overlap with the $\gamma$~Dor type pulsators, while there are numerous cases of pulsating stars with hybrid DS-$\gamma$~Dor behaviour.

\section{Summary and conclusions}
\label{Con}

This work contains the demographics and some statistical results on the largest catalogue of DS in binary systems to date, which contains 1048 DS in 1043 systems. The online form of the catalogue is given at \url{https://alexiosliakos.weebly.com/catalogue.html}. The sample was divided according to the Roche geometry of the host systems of the DS. The vast majority of the binary DS are Main-Sequence stars and they are located within the classical instability strip. The correlation between $P_{\rm orb}$-$P_{\rm pul}$ has been well established for both detached and semidetached systems, while the limit of 12.5-13~d in $P_{\rm orb}$ beyond that these quantities are uncorrelated has been constrained with three different methods.

The DS in binary systems have masses between 1.25-2.9~M$_{\odot}$, radii between 1.3-4.4~R$_{\odot}$, and temperatures between 6700-9800~K. Their dominant pulsation frequencies range between 3.5-77~d$^{-1}$, but those in detached systems show a distribution peak within the range 5-10~d$^{-1}$, while those in semidetached binaries have a peak between 15-20~d$^{-1}$. There is correlation between evolutionary stage and the dominant frequency of the binary DS that is similar between the members of short-period detached and semidetached binaries but totally different from that of single DS. In general, the younger the star the highest the dominant pulsation frequency. Moreover, we still argue that the existence of the companion star plays significant role in the earlier initiation of the pulsations and their longer preservation in comparison with the single DS stars.

Although the sample is now relatively large, the sub-sample of the binary DS with accurately determined absolute properties is still rather small (66 cases). Therefore, we strongly encourage the community to obtain radial velocities measurements of the more than 400 eclipsing systems of the catalogue that lack information on their absolute parameters. This will lead to the sample increase, which in turn, will lead us to reach the ultimate goals that are the: i)~setting of further constrains on the evolution of these stars and ii)~checking of the influence of the proximity effects on their pulsational behaviour.

\acknowledgements
The author acknowledges financial support from the NOA’s internal fellowship `SPECIES' (No.~5094). This study is partially based on observations made with the 1.2~m Kryoneri telescope, Corinthia, Greece, which is operated by the Institute for Astronomy, Astrophysics, Space Applications and Remote Sensing of the National Observatory of Athens, Greece.

\bibliography{C07references.bib}

\end{document}